\documentstyle[epsfig]{aprim10}
\input{epsf}

\newif\ifAMStwofonts

\ifoldfss
  \ifCUPmtlplainloaded \else
    \NewTextAlphabet{textbfit} {cmbxti10} {}
    \NewTextAlphabet{textbfss} {cmssbx10} {}
    \NewMathAlphabet{mathbfit} {cmbxti10} {} 
    \NewMathAlphabet{mathbfss} {cmssbx10} {} 
  \fi
  \ifAMStwofonts
    \ifCUPmtlplainloaded \else
      \NewSymbolFont{upmath} {eurm10}
      \NewSymbolFont{AMSa} {msam10}
      \NewMathSymbol{\upi}     {0}{upmath}{19}
      \NewMathSymbol{\umu}     {0}{upmath}{16}
      \NewMathSymbol{\upartial}{0}{upmath}{40}
      \NewMathSymbol{\leqslant}{3}{AMSa}{36}
      \NewMathSymbol{\geqslant}{3}{AMSa}{3E}

    \fi
  \fi
\fi 

\ifnfssone
  \newmathalphabet{\mathit}
  \addtoversion{normal}{\mathit}{cmr}{m}{it}
  \addtoversion{bold}{\mathit}{cmr}{bx}{it}
  \newmathalphabet{\mathbfit} 
  \addtoversion{normal}{\mathbfit}{cmr}{bx}{it}
  \addtoversion{bold}{\mathbfit}{cmr}{bx}{it}
  \newmathalphabet{\mathbfss} 
  \addtoversion{normal}{\mathbfss}{cmss}{bx}{n}
  \addtoversion{bold}{\mathbfss}{cmss}{bx}{n}
  \ifAMStwofonts
    \ifCUPmtlplainloaded \else
      \UseAMStwoboldmath
      \makeatletter
      \new@mathgroup\upmath@group
      \define@mathgroup\mv@normal\upmath@group{eur}{m}{n}
      \define@mathgroup\mv@bold\upmath@group{eur}{b}{n}
      \edef\UPM{\hexnumber\upmath@group}
      \new@mathgroup\amsa@group
      \define@mathgroup\mv@normal\amsa@group{msa}{m}{n}
      \define@mathgroup\mv@bold\amsa@group{msa}{m}{n}
      \edef\AMSa{\hexnumber\amsa@group}
      \makeatother
      \mathchardef\upi="0\UPM19
      \mathchardef\umu="0\UPM16
      \mathchardef\upartial="0\UPM40
      \mathchardef\leqslant="3\AMSa36
      \mathchardef\geqslant="3\AMSa3E
    \fi
  \fi
\fi 

\ifnfsstwo
  \DeclareMathAlphabet{\mathbfit}{OT1}{cmr}{bx}{it}
  \SetMathAlphabet\mathbfit{bold}{OT1}{cmr}{bx}{it}
  \DeclareMathAlphabet{\mathbfss}{OT1}{cmss}{bx}{n}
  \SetMathAlphabet\mathbfss{bold}{OT1}{cmss}{bx}{n}
  \ifAMStwofonts
    \ifCUPmtlplainloaded \else
      \DeclareSymbolFont{UPM}{U}{eur}{m}{n}
      \SetSymbolFont{UPM}{bold}{U}{eur}{b}{n}
      \DeclareSymbolFont{AMSa}{U}{msa}{m}{n}
      \DeclareMathSymbol{\upi}{0}{UPM}{"19}
      \DeclareMathSymbol{\umu}{0}{UPM}{"16}
      \DeclareMathSymbol{\upartial}{0}{UPM}{"40}
      \DeclareMathSymbol{\leqslant}{3}{AMSa}{"36}
      \DeclareMathSymbol{\geqslant}{3}{AMSa}{"3E}
    \fi
  \fi
\fi 

\ifCUPmtlplainloaded \else
  \ifAMStwofonts \else 
    \def\upi{\pi}
    \def\umu{\mu}
    \def\upartial{\partial}
  \fi
\fi

\title[Evidence for IMLR]
{ Evidence for an InterMediate Line Region in AGN's Inner Torus
Region and Its Evolution from Narrow to Broad Line Seyfert I
Galaxies }

\author[Zhu,
Zhang \& Tang]
       {Ling Zhu$^{1}$, Shuang Nan Zhang$^{1,2,3}$, and Sumin Tang$^{4}$\\
        $^1$Department of Physics and Tsinghua Center for
Astrophysics, Tsinghua University, Beijing 100084, China;\\
zhul04@mails.tsinghua.edu.cn, zhangsn@tsinghua.edu.cn \\
        $^2$Key Laboratory of Particle Astrophysics, Institute
of High Energy Physics, Chinese Academy of Sciences,\\ P.O. Box
918-3, Beijing 100049, China\\
        $^3$Physics Department,
University of Alabama in Huntsville, Huntsville, AL 35899, USA\\
       $^4$Harvard-Smithsonian Center for Astrophysics, 60
Garden St, Cambridge, MA 02138, USA }
\date{}

\pagerange{\pageref{firstpage}--\pageref{lastpage}}
\pubyear{2008}

\begin{document}

\maketitle

\label{firstpage}

\begin{abstract}
We have decomposed the broad H$\alpha$, H$\beta$ and H$\gamma$ lines
of 90 Active Galactic Nuclei (AGNs) into a superposition of a very
broad and an intermediate Gaussian components (VBGC and IMGC) and
discovered that the two Gaussian components evolve with FWHM of the
whole emission lines. We suggest that the VBGC and the IMGC are
produced in different emission regions, namely, Very Broad Line
Region (VBLR) and Intermediate Line Region (IMLR). The details of
the two components of H$\alpha$, H$\beta$ and H$\gamma$ lines
indicate that the radius obtained from the emission line
reverberation mapping normally corresponds to the radius of the
VBLR, but the radius obtained from the infrared reverberation
 mapping corresponding to IMLR, i.e., the inner boundary of the dusty torus. The existence of the IMGC
 may affect the measurement of the black hole mass in AGNs.
Therefore, the deviation of NLS1s from the M-sigma relation may be
explained naturally in this way. The evolution of the two emission
line regions may be related to the evolutionary stages of the broad
line regions of AGNs from NLS1s to BLS1s. Other evidences for the
existence of the IMLR are also presented.
\end{abstract}

\begin{keywords}
line: profiles; quasars: emission lines; galaxies: structure;
Galaxy: evolution; galaxies: active; galaxies: nuclei
\end{keywords}

\section{Introduction}

Type I AGNs are often classified into two subclasses according to
the Full Width at Half Maximum (FWHM) of their broad H$\beta$ lines,
i.e NLS1 and BLS1. There is a tight correlation between the black
hole mass and the stellar velocity dispersion of the bulge of normal
galaxies, the so called M-sigma relation (e.g. Magorrian et al.
1998; Tremaine et al. 2002;). BLS1s are found also to follow this
relation well (Greene \& Ho 2005). However, NLS1s seem to deviate
from this relation; they seem to have much smaller masses or much
higher stellar velocity dispersions (e.g., Wang \& Lu  2001; Zhou et
al. 2006). {\it We call this under-massive black hole problem of
NLS1s.} It is not clear if the luminosity dependence of dust torus
and broad line region is simultaneously related to the under-massive
black hole problem and the receding torus model that used to explain
the luminosity function of AGN (Simpson 2005). Progress towards
solving these problems may shed new lights to the understanding the
hierarchical evolution of AGN and its host galaxy, as well as the
feeding, fueling and growth of supermassive black holes. In this
paper we attempt to address both problems in a coherent way by
studying a sample of SDSS AGNs with strong broad emission lines.

\section{Line decomposition and statistical analysis}

The sample contains 90 objects with clear H$\alpha$, H$\beta$, and
H$\gamma$ line profiles that can be decomposed. They are selected by
La Mura et al (2004) from SDSS 3 based on their balmer line
intensities. 21 of them are classified as NLS1s. A double Gaussian
component model, i.e IMGC and VBGC can fit the broad line very well.
H$\alpha$, H$\beta$ and H$\gamma$ lines are fitted with their FWHM
fixed with each other; a deviation of 10\% are allowed. Several
decomposition samples are presented in Fig.1. The statistical
analysis are presented in Fig.2. H$\alpha$, H$\beta$ and H$\gamma$
lines behave similarly. Our analysis is focused on H$\beta$. The
FWHM ratios range from 1.5 to 4 in the NLS1 group which lie on the
left of Fig.2(a1). The intensity ratio of the very broad Gaussian
component to the whole line is about 0.6 in these objects. This
means that for NLS1s the intermediate Gaussian component has an
intensity comparable to the very broad one, and consequently the
FWHM of the whole line of NLS1s is dominated by the intermediate
Gaussian component. With FWHM increasing, the intermediate Gaussian
component becomes weaker and finally disappears, which can be seen
from Fig.2(b1). The whole line can be simply described by a single
very broad Gaussian component when FWHM reaches about 5000
km$\cdot$s$^{-1}$. In Fig.2(c1), FWHM ratio of the intermediate
Gaussian component to the very broad Gaussian component becomes
larger with increasing FWHM.

\begin{figure}
 \centerline{{\epsfxsize=9cm \epsffile{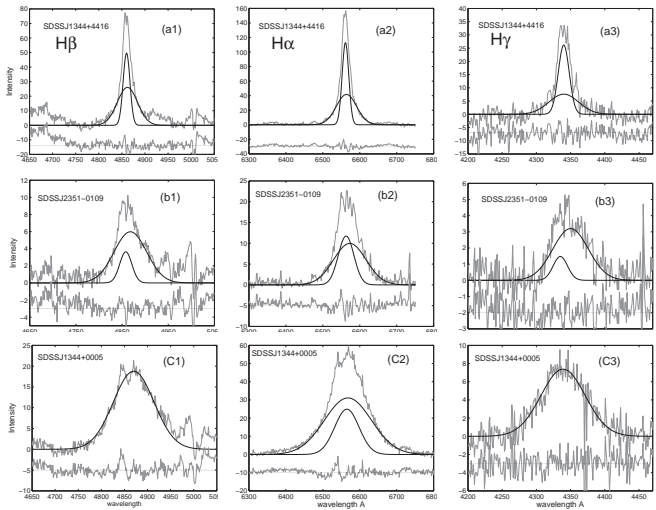}}}
    \caption[]
      {Decomposition example of broad H$\beta$, H$\alpha$ and H$\gamma$ lines. From (a) to (c), FWHM of the line
  increases, and the IMGC becomes weaker and may disappear sometimes, e.g., in subpanels c1 \& c3.
  Note that for in subpanel c2, the IMGC of H$\alpha$ line is still detectable, in contrast to
  H$\beta$ and H$\gamma$ lines, because the intermediate component of H$\alpha$ is normally much stronger than
  H$\beta$ and H$\gamma$ lines.
      }
 \label{f1}
\end{figure}

\begin{figure}
 \centerline{{\epsfxsize=9cm \epsffile{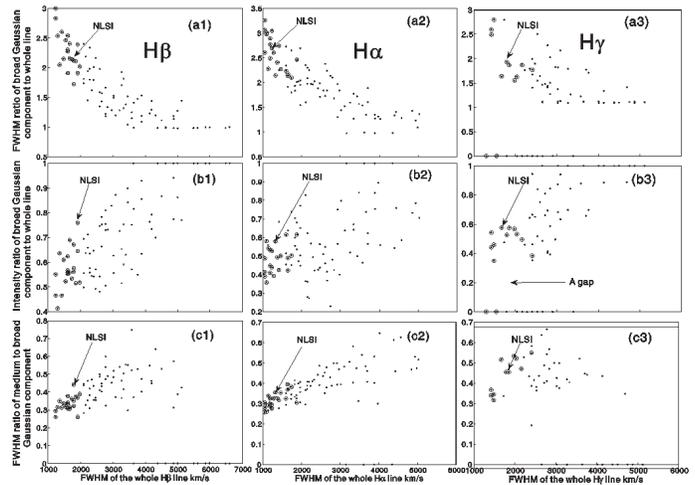}}}
    \caption[]
      {Statistical analysis of broad H$\beta$, H$\alpha$ and H$\gamma$ lines. With the
   FWHM increasing, FWHM ratio of VBGC
    to the whole line becomes smaller and finally reaches
    unity, the intensity ratio of the VBGC
    to the whole line becomes larger and finally also reaches unity, and
  the IMGC becomes broader and weaker.
      }
 \label{f2}
\end{figure}

\section{BHM correction}

Reverberation-based black hole mass is calculated by the virial
equation
\begin{equation}
M_{\rm{
BH}}=\frac{R_{\rm{ BLR}}}{G}f^2\rm{FWHM}_{\rm{ H}\beta}^2,\\
\end{equation}
where FWHM$_{\rm{ H}\beta}$ is the FWHM of the whole H${\beta}$,
which represents the virial velocity of the BLR. For an isotropic
velocity distribution, as generally assumed, $f=\sqrt{3}/2$.
$R_{\rm{ BLR}}$ is the BLR radius obtained by reverberation mapping,
which is related to an AGN's continuum luminosity by the empirical
Radius-Luminosity relation (Bentz et al. 2006):
\begin{equation}
\log{\frac{R_{\rm{ BLR}}}{\rm{lt-days}}}=(0.518 \pm 0.039)
\log{\frac{\lambda L_{\lambda}(5100 \AA)}{\rm{erg}\cdot
\rm{s}^{-1}}}-21.2\pm 1.7.
\end{equation}
This relation has been starlight-corrected. We use this equation to
calculate $R_{\rm{ BLR}}$ throughout this work. Because
reverberation mapping is used to measure the radius of the broad
line region, naturally it normally measures the radius of the inner
most emission line region, i.e. the VBLR in our analysis.
Consequently, FWHM of the very broad Gaussian component, instead of
FWHM of the whole line, should be used to calculate the black hole
mass. We therefore correct the black hole mass in this way:
\begin{equation}M_{\rm{
BHb}}=\frac{R_{\rm{
VBLR}}}{G}f^2\rm{FWHM}^2(\frac{\rm{FWHMb}}{\rm{FWHM}})^2,
\end{equation}
where FWHMb is the FWHM of the very broad Gaussian component, and
$R_{\rm{ VBLR}}$ is taken as the $R_{\rm{ BLR}}$ in the R-L
relation. It gives a more significant mass correction for NLS1s than
for BLS1s as shown in Fig.3. After such correction, NLS1s still have
smaller black hole masses but normal accretion rate in units of the
Eddington rate.
\begin{figure} 
 \centerline{{\epsfxsize=9cm \epsffile{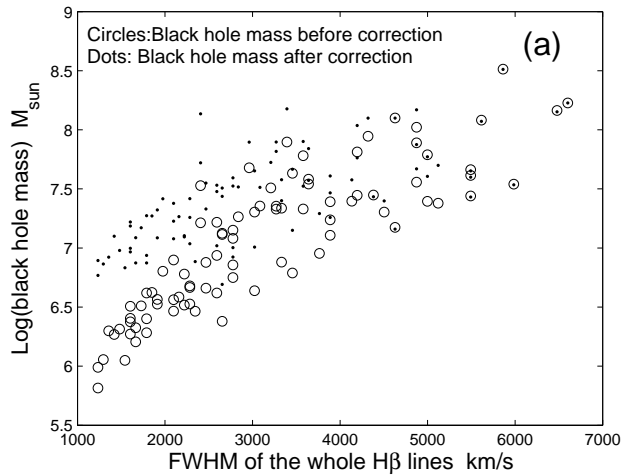}}}
   \caption[]
   {
  Black hole
  masses (in units of solar mass $M_{\sun}$) are plotted in logarithm scale. Circles are black hole
  mass calculated with equation (1). Dots are that after correction
  using equation (3). Clearly the correction is more effective for AGNs with smaller FWHM, i.e., NLS1s.
   }
\label{f3}
\end{figure}
Fifteen objects in our samples have velocity dispersion measurement
data (sigma) in Shen et al (2008). The masses of all NSL1 are well
below, but become more very close to, the predictions of the M-sigma
relation before and after the correction.The effect of the
correction is obvious, albeit small number statistics due to the
limited sample.

\section{The location of the IMLR and the evidences for its existence}

\subsection{Broad Line Region evolution}
We assume that both of the two regions (VBLR and IMLR) are bounded
by the central black hole's gravity. Then, we have
\begin{equation}
f_1^2\frac{V_{\rm{ VBLR}}^{2}}{R_{\rm{ VBLR}}}=\frac{GM_{\rm{
BH}}}{R_{\rm{ VBLR}}^{2}},
\end{equation}
and
\begin{equation}
f_2^2\frac{V_{\rm{ IMLR}}^{2}}{R_{\rm{ IMLR}}}=\frac{GM_{\rm{
BH}}}{R_{\rm{ IMLR}}^{2}}.
\end{equation}
These two equations lead
to
\begin{equation}
R_{\rm{ IMLR}}=(\frac{f_1V_{\rm{ VBLR}}}{f_2V_{\rm{
IMLR}}})^{2}R_{\rm{ VBLR}}=C_0(\frac{V_{\rm{ VBLR}}}{V_{\rm{
IMLR}}})^{2}R_{\rm{ VBLR}},
\end{equation}
where $\rm{C}_0=(\frac{f_1}{f_2})^2.$ $\rm{C}_0$ is a constant which
need to be determined by further studies. The radius measured from
reverberation is taken as $R_{\rm{ VBLR}}$. Therefore, we can
calculate the radius of IMLR from the above equation.
$R_{\rm{IMLR}}$ is proportional to $\rm{C}_0$. $\rm{C}_0=1$ is used
for all the calculation below. $R_{\rm{ IMLR}}$ is obtained from
H$\alpha$ and H$\beta$ lines separately. Fig.4 is the evolution of
$R_{\rm{ VBLR}}$ and $R_{\rm{ IMLR}}$ with luminosity. The radius of
the IMLR increases slower than VBLR with luminosity and black hole
mass increasing. The two emission regions have a trend to merge into
one with larger higher luminosity and black hole mass. as
$R_{\rm{IMLR}}\propto \rm{L5100}^{0.37\pm0.06}$ while
$R_{\rm{VBLR}}\propto \rm{L5100}^{0.52\pm0.04}$.
\begin{figure} 
 \centerline{{\epsfxsize=9cm \epsffile{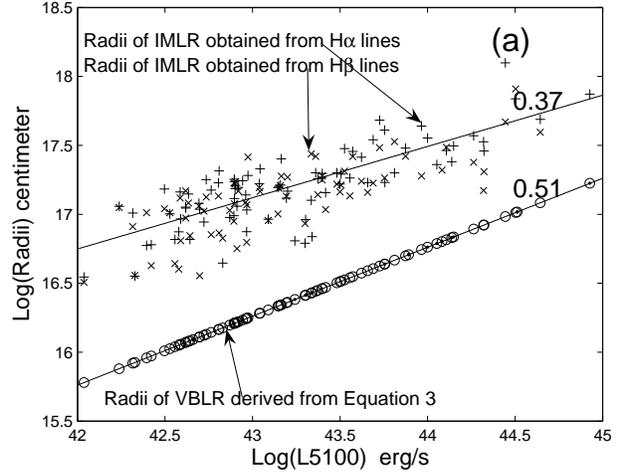}}}
   \caption[]
   {
  Correlation of the radii of IMLR and VBLR with luminosity. Circles are the VBLR radius, $\times$ represent IMLR radius derived
  from H$\beta$ lines and $+$ from H$\alpha$ lines. Circles filled with dots
  represent the objects missing the intermediate Gaussian component. The radius of the VBLR is obtained from the $R_{\rm{ BLR}} \sim$L5100
   (equation 2 not equation 3). These correlations supports a scenario of hierarchical evolution
of AGNs.
   }
\label{f4}
\end{figure}
\subsection{The location of IMLR}
The relation $R_{\rm{IMLR}}\propto \rm{L5100}^{0.37\pm0.06}$ is
consistent with the receding velocity of torus based on infrared
reverberation mapping (Suganuma et al. 2006) and the receding
velocity of torus that we can calculate based on analysis of type I
AGN fraction (Simpson 2005). Our other analysis also shows that the
IMLR may have higher density, more dusty than the VBGC, and the IMGC
show slightly Baldwin effect while the VBGC does not. The Baldwin
effect can be explained as a flattened geometry. These results
suggest that IMLR is the inner hot skin of torus.
\begin{figure} 
 \centerline{{\epsfxsize=9cm \epsffile{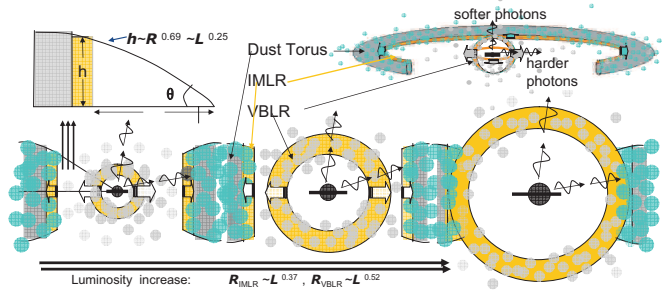}}}
   \caption[]
   {
  Cartoon of the Broad Line Region evolution. Continuum photons from
  the central region have slightly
  higher energies along the nearly edge-on direction. With black hole mass and luminosity increasing, both the
  VBLR and IMLR expand. The radius of VBLR increases faster, so the two regions have a trend to merge into
  one.
   }
\label{f5}
\end{figure}
A simple scenario of the BLR (BLR=VBLR+IMLR) evolution can be
constructed as shown in Fig.5. The inner spherical region is the
VBLR (Very Broad Line Region). It expands to a larger radius with
luminosity increase and perhaps also black hole mass increase. IMLR
is the inner part of the torus, which can be sublimated by the
central radiation and thus its radius also increases with luminosity
increase. When the luminosity is high enough, the broad line region
and torus become one single entity, but with different physical
conditions. This scenario is consistent with the dust bound BLR
hypothesis which also showed that the delay time of infrared
emission is always longer than the delay time of emission lines of
the corresponding AGN, but sometimes they are almost the same (Laor 2004; Elitzur 2006). Suganuma et al. (2006).\\

\subsection{Evidence for the existence of IMLR}

The first concrete evidence comes from the the study of the
micro-lensing of the Broad Line Region in the lensed quasar
J1131-1231 (Sluse et al. 2007, 2008). In this work they found
evidence that the H$\beta$ emission line (as well as H$\alpha$) is
differentially microlensed, with the broadest component (FWHM
$~$4000 km/s) being much more micro-lensed than the narrower
component (FWHM $~$2000 km/s). Because the amplitude of
micro-lensing depends on the size of the emitting region, this is a
clear evidence that the broadest component of the emission line is
more compact than the narrower component.

Then, we turn to the reverberation mapping experiment. If the IMLR
is really a discrete emission region apart from the VBLR, another
peak corresponding to the radius of this region might appear in the
cross correlation function between the lightcurves of the emission
line and the continuum. In fact, a possible example exists in the
database of reverberation mapping observations (Peterson et al.
1998; 2004). Mrk 79 shows obviously double peaks in the cross
correlation centroid distribution which has not been explained well
so far which can be well explained in our model (please refer to Zhu
et al. 2008 for details).

Another fact known to us is that the RMS spectrum of the broad
emission line is normally (but not always) narrower systematically
than the mean spectrum (Collin et al. 2006). If the emission lines
are produced from two distinct regions, VBLR and IMLR, they will
respond to the central continuum radiation differently. Because the
IMLR has a flattened geometry, it can create a narrower response
function to a delta impulse even with radius much larger than the
VBLR, which is confirmed by our calculations shown in Fig.6.

\begin{figure} 
 \centerline{{\epsfxsize=9cm \epsffile{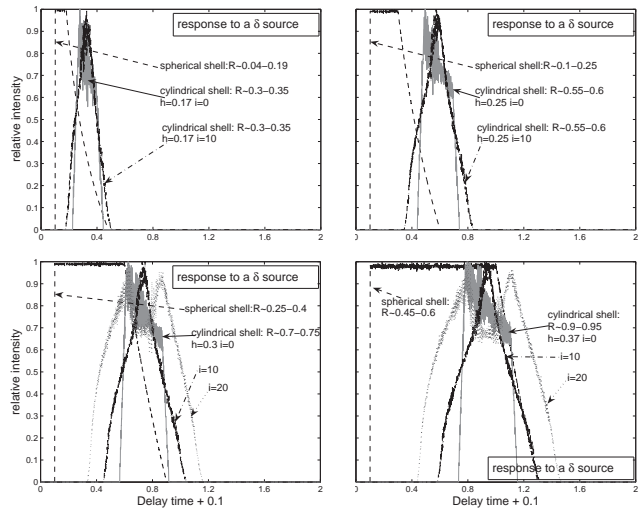}}}
   \caption[]
   {Different responses of different geometries to a delta impulse. The VBLR is assumed to be a spherical shell of uniform density, and
   the IMLR is assumed to be a cylindrical shell of uniform density. Time is normalized to arbitrary unit.
   Each figure shows a different $R_{\rm IMLR}-R_{\rm VBLR}$ pair.
   Clearly in all cases the responses of IMLR have shorter
   variability and thus will result in narrower RMS spectra than that from VBLR.
   }
\label{f6}
\end{figure}

\section{summary}

We conclude that the broad H$\beta$, H$\alpha$, and H$\gamma$ line
of the AGNs in our sample can be well fitted by a double Gaussian
model, i.e., VBGC and IMGC. The properties of the two components
indicate that the two Gaussian components originate from two
emission regions, i.e. VBLR and IMLR. The physical properties of
IMLR suggest it should be the torus inner boundary region. Based on
our model, The black hole mass measured by RM should be corrected,
after correction the NLS1 follows M-sigma relation as well as BLS1.
Therefore the problem of under-massive black hole in NLS1s appears
to have disappeared. Our results offer strong evidence of the
evolution of the broad line region (consists of a IMLR and a VBLR)
from NLS1s to BLS1s in the hierarchical evolution scenario.

Finally, we show that the study of micro-lensing provide concrete
evidence for the existence of IMLR and Mrk 79 provides possible
evidence. Narrower RMS spectrum also support our model. The
existence of a weak intermediate Gaussian component in the broad
emission lines of many sources with reverberation mapping
measurements may cause systematic biases for the measurements of the
radius of the VBLR, It will be helpful to decompose each broad
emission line into two Gaussian components as we have done here, and
then do cross-correlation analysis between the continuum and each of
the two components, in order to measure the Radius-Luminosity
relations for the two components independently. This would provide
decisive test for our model presented in this paper, as well as
providing more accurate measurements of the masses of supermassive
black holes in AGNs.

\section*{Acknowledgment}

We are extremely grateful to G. La Mura and L.C. Popovic for sending
us the processed spectra we used in this work. SNZ thanks Jianmin
Wang for many discussions, as well as sharing their early results
which motivated us to pursue the work presented here. SNZ
acknowledges partial funding support by Directional Research Project
of the Chinese Academy of Sciences under project No. KJCX2-YW-T03
and by the National Natural Science Foundation of China under grant
Nos. 10521001, 10733010,10725313, and by 973 Program of China under
grant 2009CB824800.

\label{lastpage}

\clearpage

\end{document}